 \documentclass[iop]{emulateapj-rtx4}

\shorttitle{50 BHCs in M31}
\shortauthors{Barnard et al.}

\begin{document}


\title{50 M31 black hole candidates identified by Chandra and XMM-Newton}


\author{R. Barnard, and M. R. Garcia and F. Primini}
\affil{Harvard-Smithsonian Center for Astrophysics, Cambridge MA 02138}
\and
\author{S. S. Murray}
\affil{Johns Hopkins University, Baltimore, Maryland, and CFA}


\begin{abstract}
Over the last $\sim$5 years we have identified $\sim$35 black hole candidates (BHCs) in M31 from their X-ray spectra. Our BHCs exhibited 0.3--10 keV  spectra consistent with the X-ray binary (XB) hard state, at luminosities that are above the upper limit for neutron star (NS) XBs. When our BHC spectra were modeled with a disk blackbody + blackbody model, for comparison with bright NS XBs, we found that the BHCs inhabited a different parameter space to the NS XBs. However,  BH XBs may also exhibit a thermally dominated (TD) state  that has never been seen in NS XBs; this TD state is most often observed in X-ray transients. We examined the $\sim$50 X-ray transients in our Chandra survey of M31, and found 13 with spectra suitable for analysis. We also examined 2 BHCs outside the field of view of our survey, in the globular clusters B045 and B375. We have 42 strong BHCs, and 8 plausible BHCs that may benefit from further observation. Of our 15 BHCs in globular clusters, 12  differ from NS spectra by $>$5$\sigma$. Due to improvements in our analysis, we have upgraded 10 previously identified plausible BHCs to strong BHCs. The mean maximum duty cycle of the 33 X-ray transients within 6$'$ of M31*  was 0.13; we estimate that $>$40\% of the XBs in this region contain BH accretors. Remarkably, we estimate that BHCs contribute $>$90\% of those XBs $>$10$^{38}$ erg s$^{-1}$. 

\end{abstract}





\section{Introduction}

In recent years, we have identified a number of black hole candidates (BHCs) in M31 from their X-ray spectra from XMM-Newton or Chandra, using various techniques to exclude active galactic nuclei (AGN) that may be spectrally similar \citep{barnard08,barnard09,barnard11a,barnard11b,barnard2012c,barnard13,barnard14a,barnard14b}. We have recently discovered a method of quantifying the strength of our BHC identifications \citep{barnard13,barnard14b} that involves using a double thermal emission model (disk blackbdody + blackbody) to compare our BHC spectra with the spectra of bright Galactic neutron star (NS) X-ray binaries (XBs).

The Galactic NS low mass X-ray binaries (LMXBs) were long thought to be separated into the highly luminous Z sources and the lower-luminosity atoll sources; the Z sources were further split into those that resembled Cygnus X-2, and those that resembled Scorpius X-1 \citep{hasinger89}. \cite{muno02} showed that the two populations exhibited dramatically different variability, with Z source luminosities  varying by a factor of a few while their spectra evolved over timescales of a few days, while atoll source luminosities varied by 1--2 orders of magnitude during spectral evolution over several months. 

However, three recent Galactic X-ray transients have exhibited the full range of NS LMXB behavior, going from Cyg-like to Sco-like then atoll behavior as their luminosities decreased:   XTE J1701$-$462 \citep{homan07}, IGR J17480-2446 \citep[e.g][]{chakraborty11}, and MAXI J00556$-$332 \citep[e.g.][]{sugizaki13}.      
Therefore, it is clear that NS LMXB behavior is governed by the accretion rate (which translates into luminosity). 

\citet{lin09} examined $\sim$900 RXTE observations of XTE J1701$-$462, carefully subdividing each observation so that they could study the spectral evolution in detail. They fitted each of the thousands of spectra with a double thermal model (disk blackbody + blackbody), that they developed when they examined the spectral evolution of two transient atoll LMXBs \citep{lin07}. They found that their double thermal model was successful, except for two types of spectra: hard state spectra, which are exhibited by NS and BH LMXBs at relatively low Eddington fractions and dominated by Comptonization \citep{vdk94}, and  spectra from the Z source ``horizontal branch'' \citep{hasinger89}, which required a Comptonized component in addition to the two thermal components. \citet{lin12} also obtained very similar results from their analysis of the the Sco-like Z source GX17+2.
The work of Lin et al. (2007, 2009, 2012)  covers the full range of NS LMXB behavior, which they modeled in a consistent way. Furthermore, \citet{lin10} successfully applied their model to Beppo-SAX and Suzaku spectra of the persistently bright NS XB 4U\thinspace 1705$-$44; the energy ranges were 1--150 keV and 1.2--40 keV respectively, meaning that the usefulness of the double thermal model is not confined to the RXTE pass band. 

The hard state is observed in BH and NS LMXBs \citep{vdk94}  only  at luminosities $\la$0.1 $L_{\rm Edd}$, where $L_{\rm Edd}$ is the Eddington luminosity \citep{gladstone07,tang11}. We have identified 36 BHCs that exhibit apparent hard state spectra at 0.3--10 keV luminosities too high for NS LMXBs \citep[$\ga3\times10^{37}$ erg s$^{-1}$, see ][ and references within]{barnard13,barnard14a}. When we plotted disk blackbody temperature vs. 2.0--10 keV disk blackbody luminosity for our BHCs, we found that none of our BHCs resided in the region occupied by NS LMXBs \citep[gleaned from the analysis of thousands of spectra by][]{lin07, lin09, lin12}, although some were consistent within 3$\sigma$ \citep{barnard13,barnard14a}. We classified those  BHCs $\ge$3$\sigma$ away from the NS LMXB  region as strong BHCs, and those within 3$\sigma$ as plausible BHCs; for some BHCs, the double thermal model was unconstrained, and we labeled these as plausible BHCs also.

BH XBs also exhibit a thermally dominated state that is never seen in NS XBs \citep{done03}. The thermally dominated state is most often observed in X-ray transients, where $\sim$1 keV disk blackbody emission contributes $>$75\% of the 2--20 keV flux \citep{remillard06}. We have been monitoring the central region of M31 for the last 13 years, averaging $\sim$1 observation per month; we have a total of 175 Chandra observations including our monitoring survey, deeper observations of M31* and public data from other programs. We have identified $\sim$50 transient X-ray sources in our Chandra observations \citep{barnard14a}.

In this work we examine those M31 transients with spectra consistent with the thermally dominated state, and compare double thermal fits to these spectra with the NS LMXBs in order to expand our BHC sample. We apply an improved BHC classification method to all of our BHCs, with the hope of upgrading some of the plausible BHCs to strong BHCs. 

We also examine two BHCs in globular clusters (GCs) outside of the area monitored by our Chandra survey. The first is located in the GC B045 (also known as Bo 45), following the naming convention of the Revised Bologna Catalogue v. 3.4 \citep{galleti04,galleti06,galleti07,galleti09}. \citet{barnard08} identified the X-ray source as a BHC because its hard spectrum and high variability indicated that it was in the canonical BH hard state, but the 0.3--10 keV luminosity exceeded the Eddington limit for a 1.4 M$_\odot$ NS. The second BHC resides in the GC B375 (Bo 375), and was identified as a BHC by \citet{distefano02}. In \citet{barnard08} we mistakenly said that the spectrum (well described by a 0.90$\pm$0.10 keV blackbody and a power law with photon index 1.73$\pm$0.18) was typical of a bright NS XB; however, bright NS XBs fitted with such models yield considerably higher blackbody temperatures \citep[e.g. $\sim$2 keV for Sco X-1,][]{barnard03}. 

\section{Observations and data analysis}


\begin{table*}[!t]
\begin{center}
\caption{List of properties for our Chandra BHCs. First we give identifications in previous papers and angular distance from M31*, followed by maximum and minimum 0.3--10 keV luminosity along with the best fit $\Gamma$  for those observations ($^a$ indicates mean $\Gamma$ for all spectra of that source with $>$200 counts). We then give the number of outbursts  for transients (P for persistent), then present our two estimates of the duty cycle. Finally we indicate variability with $\chi^2$/dof for constant luminosity. Numbers in parentheses indicate 1$\sigma$ uncertainties on the last digit. Upper limits to luminosities are  quoted at the 3$\sigma$ level.} \label{chantab}
\renewcommand{\tabcolsep}{5pt}
\renewcommand{\arraystretch}{1.}
\begin{tabular}{cccccccccccccc}
\tableline\tableline
Src & ID & $D_{\rm M31*}$/ $'$ & $L_{\rm Max}^{37}$  & $\Gamma_{\rm Max}$ & $L_{\rm Min}^{37}$  & $\Gamma_{\rm Min}$ & $N_{\rm O}$ & DC 1 & DC 2 & $\chi^2_{\rm con}$/dof \\
\tableline 
S109 & BH1 & 15.85 & 37 (5) & 1.5 (3) & 14.1 (17) & 1.4 (3)$^a$ & P &  &  & 230/76 \\
S111 & T1 & 6.24 & 18.3 (17) & 2.0 (4) & $<$0.06 & 1.9 (5)$^a$ & 2 & 0.04 & 0.06 & 234/79 \\
S117 & T2 & 5.11 & 53 (3) & 4.9 (6) & $<$0.04 & 1.7 & 1 & 0.05 & 0.05 & 3957/99 \\
S122 & BH2 & 5.28 & 15 (2) & 1.47 (7)$^a$ & 1.73 (15) & 1.47 (7)$^a$ & P &  &  & 1527/164 \\
S151 & BH3 & 4.06 & 31 (2) & 2.5 (5) & 4.31 (17) & 1.552 (15)$^a$ & P &  &  & 28720/166 \\
S159 & BH4 & 4.63 & 5.5 (5) & 1.89 (7)$^a$ & 0.39 (0.10) & 1.89 (7)$^a$ & P &  &  & 3999/167 \\
S167 & BH5 & 6.85 & 7.9 (1.6) & 1.5 (1)$^a$ & 1.7 (2) & 1.5 (1)$^a$ & P &  &  & 568/157 \\
S168 & BH6 & 4.83 & 19.0 (18) & 1.58 (3)$^a$ & 0.6(2) & 1.58 (3)$^a$ & P &  &  & 5639/162 \\
S179 & BH7 & 2.49 & 21 (2) & 2.7 (5) & 4.9(13) & 1.69 (3)$^a$ & P &  &  & 669/168 \\
S199 & BH8 & 19 & 19 (3) & 1.8 (4) & $<$0.4 & 1.37 (15)$^a$ & Many & 0.33 & 0.33 & 664/26 \\
S214 & BH9 & 0.9 & 8.5 (4) & 2.36 (16)$^a$ & 0.05 & 2.36 (16)$^a$ & 3 & 0.08 & 0.09 & 3482/94 \\
S223 & BH10 & 2.78 & 4.5(4) & 1.64 (5)$^a$ & 0.55(7) & 1.64 (5)$^a$ & P &  &  & 4983/167 \\
S233 & T5 & 0.66 & 9.8(13) & 2.0 (5) & $<$0.04 & 1.7 & 2 & 0.02 & 0.06 & 317/100 \\
S236 & BH11 & 0.41 & 23(2) & 1.41 (7)$^a$ & $<$0.05 & 1.41 (7)$^a$ & 1 (turn off) & 0.16 & 0.12 & 2319/95 \\
S251 & U2 & 9.2 & 320 (8) & 3.9 (5) & $<$0.4 & 1.7 & 1 & 0.11 & 0.08 & 61651/69 \\
S265 & BH12 & 4.52 & 9.0(18) & 2.5 (9) & 1.5(2) & 2.08 (4) & P &  &  & 3701/167 \\
S269 & BH13 & 0.26 & 4.9(3) & 1.78 (5)$^a$ & 0.76(8) & 1.78 (5)$^a$ & P  &  &  & 3625/170 \\
S276 & BH14 & 5.55 & 14 (2) & 2.9 (7) & $<$0.05 & 2.6 (3)$^a$ & 1 & 0.30 & 0.17 & 8381/98 \\
S286 & BH15 & 0.5 & 7.9 (9) & 1.61 (4)$^a$ & 1.2 (2) & 1.61 (4)$^a$ & P  &  &  & 3472/170 \\
S287 &  & 2.1 & 20.5 (5) & 3.67 (13) & 0.07(2) & 1.7 & 1 & 0.05 & 0.03 & 82/92 \\
S289 & BH16 & 0.62 & 26.3 (6) & 1.58 (3)$^a$ & 0.6 (2) & 1.58 (3)$^a$ & P  &  &  & 16817/164 \\
S293 & B128  & 4.96 & 5.9 (5) & 1.64 (10)$^a$ & $<$0.03 & 1.64 (10)$^a$ & 2 & 0.14 & 0.13 & 2348/87 \\
S297 & BH17 & 0.89 & 8.0 (3) & 1.91 (4)$^a$ & 0.73 (14) & 1.91 (4)$^a$ & P  &  &  & 4602/170 \\
S299 & BH18 & 1.12 & 20 (2) & 1.50 (2)$^2$ & 6.8 (8) & 1.8 (3) & P &  &  & 842/170 \\
S300 & BH19 & 9.26 & 21 (3) & 1.9 (6) & 0.75(18) & 1.84 (5)$^a$ & P  &  &  & 9096/127 \\
S322 &  & 1.62 & 13 (2) & 2.5 (6) & $<$0.04 & 1.7 & 1 & 0.02 & 0.04 & 227/72 \\
S327 & BH20 & 15.1 & 62 (3) & 1.14 (14) & 30 (2) & 1.89 (2)$^a$ & P  &  &  & 587/73 \\
S330 & T8 &8.4  &2.7 (4)& 2.10 (17)$^a$  &  $<$0.06 & 2.10 (17)$^a$ & 1 (turn on) & 0.73 & 0.79 & 2941/158  \\
S331 & T13 & 1.6 & 6.1 (6) & 4.02 (17) & $<$0.0016 & 1.7 & 1 & 0.018 & 0.05 & 171/45 \\
S335 & BH21 & 3.2 & 20.6(17) & 1.9 (4) & 5.6 (4) & 1.7 & P  &  &  & 441/108 \\
S339 & T9 / U1 & 2.4 & 394 (2) & ? & $<$0.025 & 1.74 (2)$^a$ & 1 & 0.06 & 0.04 & 37255/89 \\
S345 & BH22 & 2 & 23 (2) & 1.70 (5)$^a$ & 0.82 (18) & 1.7 & P &  &  & 5782/108 \\
S353 &  & 3.5 & 5.8 (15) & 1.6 (3)$^a$ & $<$0.05  & 1.70 (5)$^a$ & 5 & 0.20 & 0.21 & 1984/112 \\
S358 & BH23 & 5.7 & 8.9(13) & 2.5 (6) & $<$0.18 & 1.6 (3)$^a$ & P  &  &  & 651/167 \\
S365 &  & 2.8 & 14(2) & 2.9 (9) & $<$0.009 & 1.78 (4) & 1 & 0.06 & 0.04 & 503/60 \\
S372 & BH24 & 4.3 & 7.2(13) & 1.8 (4) & 2.8(3) & 2.37 (19)$^a$ & P &  &  & 801/168 \\
S373 & BH25 & 2.9 & 7.2 (11) & 1.9 (6) & 3.1(9) & 1.78 (8) & P &  &  & 406/107 \\
S386 & BH26 & 3.6 & 8.2 (13) & 1.84 (5)$^a$ & 1.3 (2) & 1.4 (3)  & P &  &  & 5245/170 \\
S389 & BH27 & 3.6 & 13.0 (18) & 2.01 (5)$^a$ & 0.37 (9) & 1.84 (5)$^a$ & P &  &  & 14549/170 \\
S391 & BH28 & 4.2 & 7.7 (3) & 1.69 (4)$^a$ & 1.8 (2)  & 2.1 (5) & P &  &  & 2885/169 \\
S396 & BH29 & 4.1 & 46.8 (5) & 1.46 (8)$^a$ & $<$0.07 & 1.69 (4)$^a$ & 1 & 0.05 & 0.05 & 20445/99 \\
S411 & BH30 & 5.6 & 12 (2) & 1.9 (6) & $<$0.16 & 1.46 (8)$^a$ & Many & 0.26 & 0.42 & 2667/145 \\
S415 & BH31 & 5.1 & 20 (3) & 1.6 (3)  & 7.5 (14) & 1.9 (2)$^a$ & P  &  &  & 511/102 \\
S438 & BH32 & 13.2 & 14.2 (12) & 1.6 (2) & $<$0.007 & 1.47 (2)$^a$ & 3 & 0.70 & 0.13 & 3195/76 \\
S448 &  & 6.9 & 97 (6) & 3.8 (4) & $<$0.04 & 1.58 (2)$^a$ & 2 & 0.06 & 0.10 & 3198/89 \\
S484 & BH33 & 9.8 & 10 (3) & 1.94 (6)$^a$ & 1.4 (2) & 1.7 & P &  &  & 1450/111 \\
S487 & BH34 & 10.1 & 10.2 (13) & 1.9 (5) & 3.3 (11) & 1.94 (6)$^a$ & P &  &  & 100/39 \\
S497 & BH35 & 13.9 & 12.8 (15) & 1.7 (4) & 1.07 (16) & 1.49 (5)$^a$ & P  &  &  & 2578/78 \\

\tableline
\end{tabular}

\end{center}
\end{table*}


\begin{table*}[!t]
\begin{center}
\caption{ For each BHC we give the observation number, degree of pile-up, then the difference in $\chi^2$ between different emission models, i.e  (H)ard, (T)hermal, and (S)teep power law: $\Delta1$ = $\chi^2_{\rm T}-\chi^2_{\rm H}$ and $\Delta2$ = $\chi^2_{\rm H}-\chi^2_{\rm S}$; U indicates that the steep power law model was unconfined. We then give the possible spectral states (probability $>$0.05), with boldface indicating the preferred state; ``H (dp)'' indicates a hard state where both components are detected; we note that the spectrum for S339 was piled up, and we quote the spectral results found by  Nooraee et al. (2012). Finally we give the spectral parameters: absorption; disk blackbody temperature and luminosity if applicable; photon index and power law luminosity if applicable; $\chi^2$/dof.
} \label{spectab}
\renewcommand{\tabcolsep}{5pt}
\renewcommand{\arraystretch}{1.}
\begin{tabular}{cccccccccccccc}
\tableline\tableline
BH & Obs & Pileup& $\Delta1$ &  $\Delta2$ & States &  $N_{\rm H}^{21}$  & $kT$ / keV & $L_{\rm D}^{37}$ & $\Gamma$ & $L_{\rm P}^{37}$ & $\chi^2$/dof \\
\tableline
B045 & 0402560901 &  & 224 &  43 & H {\bf S} &  2.3 (4) & 2.58 (14) & 13.0 (16) & 2.6 (4) & 11.2 (14) & 619/609 \\
B375 & 0402561201 &  & 319 &  83 & {\bf S} &  1.46 (14) & 1.69 (13) & 24 (4) & 2.01 (15) & 33 (3) & 742/748 \\
S109 & 303 &  0.2\% & 4 &  $\dots$ & {\bf H} T U &  5.9 (9) & $\dots$ & $\dots$ & 1.86 (16) & 18.9 (14) & 38/50 \\
S111 & 7139 & 3.9\% &  $-$2 &  $\dots$ & H {\bf T} U &  2.3 (13) & 1.4 (2) & 10.2 (9) & $\dots$ & $\dots$ & 10/15 \\
S117 & 7064 & 1.3\% & $-$29 &  $\dots$ & T U &  0.07 & 3.35 (15) & 3.04 (18) & $\dots$ & $\dots$ & 60/23 \\
S122 & 303 &3.0\% &  0 &  $\dots$ & {\bf H} T U &  1.5 (6) & $\dots$ & $\dots$ & 1.52 (14) & 6.8 (4) & 33/30 \\
S151 & 13827 & 6.4\% & 76 &  13 & H {\bf S} &  0.7 & 1.99 (16) & 10.2 (6) & 3.0 (5) & 3.9 (6) & 160/181 \\
S159 & 0690600401 & & 13 &  $\dots$ & {\bf H} T U &  3.8 (5) & $\dots$ & $\dots$ & 1.78 (10) & 5.3 (3) & 45/78 \\
S167 & 14196 & 0.5\%& 10 &  $\dots$ & {\bf H} T U &  1.9 (6) & $\dots$ & $\dots$ & 1.52 (11) & 4.03 (18) & 54/53 \\
S168 & 0112570101&  & 397 &  5 & {\bf H} S &  1.12 (8) & $\dots$ & $\dots$ & 1.73 (3) & 7.02 (11) & 429/411 \\
S179 & 0112570101&  & 793 &  22 & {\bf H} (dp) &  2.1 (4) & 0.161 (18) & 1.0(7) E$-$3 & 1.76 (4) & 6.6 (2) & 519/493 \\
S199 & 10715 & 0.2\% & 2 &  $\dots$ & {\bf H} T U &  0.7 & $\dots$ & $\dots$ & 1.17 (12) & 13.7 (13) & 15/14 \\
S214 & 1575 & 0.6\% & 76 &  11 & {\bf  H} S &  1.02 (16) & $\dots$ & $\dots$ & 1.75 (6) & 5.29 (14) & 106/109 \\
S223 & 13827 & 4.8\% &  15 &  $\dots$ & {\bf H} U &  1.3 (3) & $\dots$ & $\dots$ & 1.57 (9) & 3.73 (13) & 72/67 \\
S233 & 4719 & 11\% &  $-$1 &  $\dots$ & H {\bf T} U &  0.7 & 1.14 (14) & 5.6 (5) & $\dots$ & $\dots$ & 4/12 \\
S236 & 303 & 13\% &  12 &  $\dots$ & {\bf H} T U &  1.0 (5) & $\dots$ & $\dots$ & 1.43 (12) & 9.8 (5) & 58/45 \\
S251 & 0690600401 & & $-$1012 &  $\dots$ & {\bf T} &  3.33 (5) & 0.572 (4) & 39.2 (4) & $\dots$ & $\dots$ & 615/570 \\
S265 & 13825 & 3.0\% & 41 &  $\dots$ & {\bf H}  U &  1.6 (2) & $\dots$ & $\dots$ & 2.01 (7) & 5.8 (2) & 90/98 \\
S269 & 14197 & 6.0\% &  21 &  $\dots$ & {\bf H} U &  1.2 (3) & $\dots$ & $\dots$ & 1.64 (9) & 3.33 (13) & 72/64 \\
S276 & 9521 & 2.5\% & 1 &  $\dots$ & H  {\bf T} U &  0.7 & 1.2 (3) & 4.8 (7) & $\dots$ & $\dots$ & 6/6 \\
S286 & 14198 & 6.7\% & 6 &  $\dots$ & {\bf H} T U &  1.3 (3) & $\dots$ & $\dots$ & 1.58 (8) & 4.01 (14) & 75/74 \\
S287 & 14196 & 12\% & $-$94 &  186 & {\bf S} &  0.7 & 4.2 (2) & 3.1 (3) & 3.13 (18) & 3.4 (4) & 111/114 \\
S289 & 13825 & 15\% & 23 &  $\dots$ & {\bf H} T U &  1.15 (8) & $\dots$ & $\dots$ & 1.42 (5) & 10.3 (2) & 161/165 \\
S293 & 14196 & 1.1\% & 20 &  $\dots$ & {\bf H} T  U &  0.8 (6) & $\dots$ & $\dots$ & 1.55 (10) & 4.9 (2) & 56/63 \\
S297 & 14197 & 7.3\% & 46 &  5 & {\bf  H} S &  1.4 (3) & $\dots$ & $\dots$ & 2.06 (9) & 3.71 (16) & 94/76 \\
S299 & 13825 & 14\% & 37 &  $\dots$ & {\bf H} T U &  0.94 (18) & $\dots$ & $\dots$ & 1.47 (5) & 9.1 (2) & 140/154 \\
S300 & 0112570101 & & 20 &  8 & {\bf H} (dp) T &  3.4 (6) & 1.3 (4) & 1.6 (7) & 1.7 (4) & 4.4 (13) & 210/226 \\
S322 & 10554 & 9.6\% & 1 &  $\dots$ & H {\bf T} U &  0.7 & 0.76 (8) & 5.9 (4) & $\dots$ & $\dots$ & 7/10 \\
S327 & 0402560901 & & 19 &  20 & {\bf H} T &  2.78 (8) & $\dots$ & $\dots$ & 1.62 (3) & 43.8 (5) & 755/735 \\
S330 & 14198 & 1.1\% & 19 &  $\dots$ & {\bf H} U &  0.7 & $\dots$ & $\dots$ & 2.04 (12) & 0.51 (4) & 9/13 \\
S331 & 0727960401 & & 13 &  $\dots$ & {\bf T} U &  0.74 (16) & 0.385 (19) & 3.8 (3) & $\dots$ & $\dots$ & 60/65 \\
S335 & 13827 & 6.8\% &  34 &  $\dots$ & {\bf H} T U &  2.2 (3) & $\dots$ & $\dots$ & 1.75 (6) & 10.7 (3) & 127/145 \\
S339 & 11278 & 30\% & $-$3 &  $\dots$ & {\bf T} Pileup &  0.7 & 0.72 (17) & 38 (9) & $\dots$ & $\dots$ & 65/49 \\
S345 & 0112570101 & & 252 &  4 & {\bf H} (dp) &  0.8 (3) & 1.2 (5) & 1.5 (7) & 1.6 (2) & 5.9 (9) & 454/398 \\
S353 & 303 & 2.8\% & $-$2 &  $\dots$ & {\bf H} T U &  2.3 (11) & $\dots$ & $\dots$ & 1.6 (2) & 3.5 (3) & 15/13 \\
S358 & 14198 & 1.5\% &  9 &  $\dots$ & {\bf H} T U &  1.3 (2) & $\dots$ & $\dots$ & 1.66 (7) & 5.32 (16) & 111/95 \\
S365 & 7068 & 11\% & $-$1 &  $\dots$ & H {\bf T} U &  0.7 & 0.85 (5) & 6.0 (3)  & $\dots$ & $\dots$ & 29/39 \\
S372 & 0112570101 & & 293 &  8 & {\bf H} S &  1.27 (2) & $\dots$ & $\dots$ & 1.901 (8) & 4.933 (15) & 265/294 \\
S373 & 0112570101 & & 119 &  9 & {\bf H} S &  1.10 (13) & $\dots$ & $\dots$ & 1.51 (4) & 4.88 (11) & 254/236 \\
S386 & 14196 & 3.3\% & 8 &  13 & H T {\bf S} &  0.7 & 1.61 (14) & 2.8 (3) & 3.3 (11) & 0.8 (3) & 95/81 \\
S389 & 14197 & 4.6\% & 17 &  $\dots$ & {\bf H} T  U &  1.5 (2) & $\dots$ & $\dots$ & 1.86 (8) & 4.47 (15) & 69/76 \\
S391 & 13826 & 0.3\% & 2 &  $\dots$ & {\bf H} T U &  2.0 (10) & $\dots$ & $\dots$ & 1.94 (17) & 6.4 (6) & 21/26 \\
S396 & 1577 &  29\% & 11 &  7 & {\bf S} &  0.7 & 2.3 (3) & 23 (2) & 3.6 (14) & 8 (3) & 72/56 \\
S411 & 13299 & 2.2\% & 2 &  $\dots$ & H {\bf T} U &  1.9 (17) & 1.5 (3) & 7.1 (8) & $\dots$ & $\dots$ & 8/9 \\
S415 & 0112570101 & & 834 &  $-$1 & {\bf H} (dp) &  0.7 & 0.43 (9) & 0.5 (2) & 1.54 (5) & 9.5 (3) & 450/487 \\
S438 & 8184 & 0.7\% & 2 &  $\dots$ & H {\bf T} U &  0.7 & 1.39 (13) & 9.5 (7) & $\dots$ & $\dots$ & 28/23 \\
S448 & 9523 & 8.5\% & 1 &  $\dots$ & H {\bf T} U &  0.7 & 0.54 (6) & 22.0 (12) & $\dots$ & $\dots$ & 28/28 \\
S484 & 13825 & 0.3\% & 20 &  $\dots$ & {\bf H} T U &  1.4 (5) & $\dots$ & $\dots$ & 1.83 (11) & 4.7 (3) & 57/61 \\
S487 & 0112570101 & & 130 &  2 & {\bf H} (dp) &  0.9 (3) & 0.9 (3) & 1.2 (7) & 1.4 (3) & 5.1 (9) & 246/228 \\
S497 & 14198 & 0.1\% & 14 &  0 & {\bf H} S  &  1.9 (5) & $\dots$ & $\dots$ & 1.77 (10) & 6.0 (3) & 85/68 \\

\tableline
\end{tabular}

\end{center}
\end{table*}

An overview of our survey of 528 M31 X-ray sources in 174 Chandra observations is presented in \citet{barnard14a}. We refer the reader to that paper for the details of the analysis. In this work, we concentrated on the 112 ACIS observations in our survey because there is no way to extract reliable spectra from the 62 HRC observations. For some BHCs, we also examined XMM-Newton observations following the procedures outlined in \citet{barnard13}. Chandra observations were analyzed with CIAO v4.5 while XMM-Newton observations were analyzed with SAS ver. 13.0; X-ray spectra were modeled with XSPEC v12.8.

The Chandra observations are susceptable to  pile-up (2 or more photons arriving in the same detection cell within a particular exposure). Piled-up events can either result in a single photon with an energy equivalent to the sum of the energies of the two real events, or the event can be rejected because it doesn't look real \citep[see e.g.][]{davis01}. To estimate the degree of pile-up, we created a natively binned image of each X-ray source with no filtration, and found the highest number of counts in a 3$\times$3 pixel area (the size of a Chandra ACIS detection cell); from this we obtained the number of counts per frame, $n$. The pileup fraction, $f_{\rm p}$, is then given by: $f_{\rm p} \simeq n/2-(1/12)*n^2$ according to ACIS documentation.

For this study, we only considered transients with $>$200 counts in their X-ray spectra for at least one observation. We refer to these X-ray sources by the source number in our catalog \citep[S1--S528, ][]{barnard14a}. We have already highlighted BHCs that appear to exhibit hard state spectra at luminosities that are too high for NS LMXBs; our new sample exhibits spectra consistent with a disk blackbody with inner disk temperature $kT_{\rm DBB}$ $\la$1 keV. 

We estimated the duty cycle of each transient in two ways. The first of these was to assess the percentage of observations where the target was detected at a $>$3$\sigma$ Significance. Since the roll angle was unconstrained, each observation only contained a subset of all the X-ray sources; hence, we only considered observations where the transient could be observed when making this estimate of the duty cycle (DC1).  The second duty cycle estimate was made by comparing the duration of the outburst with the total observing time. To do this we measured the time between the last observation before the outburst was detected at $>$3$\sigma$ to the the first observation in which the transient detection goes below 3$\sigma$; this estimate of the duty cycle (DC2) is an upper limit.

For each object in our sample, we identify the observation that provides the highest quality BHC spectrum; this can be a Chandra ACIS observation (ObsID 303--14198) or a XMM-Newton observation (ObsID 0112570101--072960401); for XMM-Newton observations, we only analyzed the pn data. We fitted a double thermal model to the best spectrum for each object (WABS*(DISKBB+BB) in XSPEC); if the fit was unconstrained, then we classified the object as a plausible BHC unless the spectrum was too soft to be a NS LMXB (e.g. with no emission above 2 keV). We also fitted more traditional models to these spectra: absorbed power law, absorbed disk blackbody, and absorbed disk black body + power law to represent the hard, thermal and steep power law states respectively \citep{remillard06}.

We estimated the uncertainties in each fit by generating 1000 spectra from the best fit model using the XSPEC command {\sc fakeit}; random variations were introduced to each simulated  spectrum that were governed by the statistical properties of the original spectrum. We found the best fit for each simulated spectrum, and ranked the values of each parameter from lowest to highest; the 1$\sigma$ uncertainties were obtained from the 16$^{\rm th}$ and 84$^{\rm th}$ percentiles.

For the best double thermal fit to each spectrum, we examined the temperature and 2--10 keV luminosity for each component. Each spectrum was assessed according to three criteria, following \citet{barnard13, barnard14b}. The minimum disk blackbody temperature, $kT_{\rm DBB}$, for NS LMXBs depends on the luminosity: 1.0 keV / 1.2 keV  for luminosities below / above 2$\times 10^{37}$ erg s$^{-1}$ respectively; the minimum NS LMXB blackbody temperature, $kT_{\rm BB}$, is $\sim$1.5 keV; finally, the disk blackbody contribution to the 2--10 keV spectrum, $f_{\rm DBB}$, is $>$45\% for NS LMXBs. For our BHC spectra, we expect $kT_{DBB}$, $kT_{\rm BB}$ and $f_{\rm DBB}$ to be substantially lower than these minima; a cooler disk blackbody naturally leads to a smaller contribution to the 2--10 keV luminosity. 

For parameters  that are below the NS minimum, we calculate the probability that the observed value is consistent with a NS LMXB: $P_{\rm DBB}$ = erfc[(1.0$-kT_{\rm DBB})/\sigma/2^{0.5}$] or  erfc[(1.2$-kT_{\rm DBB})/\sigma/2^{0.5}$] depending on $L_{\rm DBB}$ (see above); $P_{\rm BB}$ =  erfc[(1.5$-kT_{\rm BB})/\sigma/2^{0.5}$]; $P_{\rm f}$ =  erfc[(0.45$-f_{\rm DBB})/\sigma/2^{0.5}$]. The probability that the BHC is consistent with being a NS LMXB, $P_{\rm NS}$, is then $P_{\rm DBB}\times P_{\rm BB}\times P_{\rm f}$. If a parameter exceeds the NS LMXB threshold, then the probability of that parameter being consistent with a NS LMXB is 1.  We assign a Rank to the BHC based on the probability of being consistent with a NS LMXB: Rank = $-$log($P_{\rm NS}$). A Rank $>$2.6 indicates $>$3$\sigma$  deviation from a NS spectrum, while a Rank $>$6.2 indicates a $>$5$\sigma$ deviation. This approach to identifying strong BHCs is an improvement upon the one  used in \citet{barnard13}; therefore we applied this analysis to our BHCs  previously identified by their hard state spectra.

\section{Results}

\subsection{Basic properties of the BHCs}

In addition to the 35 BHCs discussed in \citet{barnard13}, we  obtained usable spectra from 13 X-ray transients from our 13 year Chandra monitoring campaign; these include 2 ultraluminous X-ray sources (ULXs) that exhibited 0.3--10 keV luminosities $>$2$\times 10^{39}$ erg s$^{-1}$ \citep{kaur12,nooraee12,middleton13,barnard13b}.   The other $\sim$40 transients in our Chandra survey had insufficient counts in their spectra. With the addition of XB045 and XB375, which lie outside the Chandra survey, our sample contains 50 BHCs. 
In Table~\ref{chantab} we present a summary of our Chandra results for 48 BHCs (XB045 and XB375 were not included in our Chandra survey); this table is described in the following three paragraphs. 

For each source (named following Barnard et al. 2014a), we provide its identity in previous papers; BH1--BH35 are BHCs that were  analyzed in Barnard et al. (2013a), while T1--T9 are transients discussed in Barnard et al. (2012); T13 was discovered later (Barnard et al. 2014b). U1 and U2 are ultraluminous transients (Barnard et al., 2012; 2013b), and B128 is a GC BHC identified in Barnard et al. (2014a). We also provide the angular distance from M31*.

 We then present the maximum and minimum 0.3--10 keV luminosity observed in our Chandra observations of that source, along with the photon index from the best fit power law model; thermally dominated spectra are indicated by $\Gamma$ $\ga$3. If a source produced less than 200 photons during either of these observations, then we assumed the mean $\Gamma$ for all observations of that source with $>$200 counts; these values are indicate by $^a$. If we were unable to fit a  hard state spectra for a given source, we assume $\Gamma$ = 1.7 for the minimum luminosity. 

Finally we present the timing properties of each source. First we give the number of outbursts, if any; persistently bright X-ray sources are indicated with ``P''. If the number of outbursts is unclear, we simply say the source has many. Next are the two estimates of the duty cycle, DC1 and DC2; these are only provided for the transients. Finally we provide the $\chi^2$/dof from best fitting a constant intensity to the lightcurve (taken from Table 1 of Barnard et al. 2014a), to indicate the level of variability.

\subsection{Fitting canonical BH models}

We summarize our modeling of the BHC spectra with canonical BH models (hard state, thermally dominated state, steep power law state, Remillard \& McClintock 2006)
 in Table~\ref{spectab}. All three states consist of thermal and Comptonized emission components; however, the hard state is dominated by the Comptonized component and may be approximated by a power law for lower quality spectra, while the thermally dominated state may be approximated by a disk blackbody; for BHs in the steep power law states, the photon index of the Comptonized component is $>$2.4 \citep{remillard06}. For each source we first give the observation that best supports our case for a BH accretor. We then compare the $\chi^2$ values for three spectral models (WABS*DISKBB; WABS*POWERLAW; WABS*[DISKBB + POWERLAW]): $\Delta1$ shows the difference in $\chi^2$ between the disk black body model and the power law model while $\Delta2$ shows the difference between the power law model and disk blackbody +  power law model.
 Next we show the possible states, with our preferred model indicated in boldface. Finally we give the best fit parameters: absorption,  temperature and luminosity of the disk blackbody component (if applicable), photon index and luminosity of the power law component (if applicable), and $\chi^2$/dof for our preferred model.

In most cases, the preferred model is the one with the lowest $\chi^2$/dof. However, if there is no significant difference between the fits, then we consider whether the BHC is persistent or transient: we favor a hard state for a persistent source, and a thermally dominated state for a transient. Furthermore, a hard state is preferred to a steep power law state if the disk temperature is higher and $\Gamma$ is lower than expected for the steep power law state. S179, S300, S345, S415, and S487 have sufficiently good spectra to constrain the thermal components in the hard state spectra. 

 We see examples of BHCs consistent with all three canonical states. S151, S287, S386 and S396 are consistent with the steep power law state, but have $\Gamma$ $>$ 3, and are therefore particularly soft; also, $\Gamma$ = 2.6$\pm$0.4 for B045, meaning that it could be very soft too. B375  appears to be rather hot and rather hard for the SPL, but are consistent within uncertainties; the simple power law model does not yield an acceptable fit.

\subsection{Fitting double thermal models}

Table~\ref{sumtab} summarizes our results. We first give the source number of each BHC in our survey paper \citep[S1--S528, ][]{barnard14a}. Then we present the temperature and 2--10 keV luminosity for the disk blackbody and blackbody components respectively, along with the fractional contribution of the disk blackbody to the 2--10 keV emission; luminosities are normalized to 10$^{37}$ erg s$^{-1}$ and assume a distance of 780 kpc \citep{stanek98}. These data are followed by the BHC Rank (i.e. $-$log($P_{\rm NS}$)), and the class of the BHC: strong (S) or plausible (P). Finally we present any comments. Globular clusters are indicated with ``G'', and the GC name in parentheses, following the Revised Bologna Catalog v. 3.4 \citep{galleti04,galleti06,galleti07,galleti09}; transients are indicated by ``T'', and include ULXs labeled ``U''; ``P$>$S'' shows that the BHC was previously classified as a plausible BHC in \citet{barnard13}; ``Soft'' indicates a spectrum with little flux above $\sim$2 keV. Sources where the disk blackbody + blackbody model was unconstrained are indicated by dots.

 We find that 42 BHCs exhibited a Rank $>$2.6, and differed from the NS LMXB spectra by $>$3$\sigma$; these are classed as strong BHCs, and include 10 systems that have been promoted from plausible BHC classification \citep{barnard13}. Previously, we considered each criterion separately, but now we combine the probabilities for each criterion into one. Furthermore, 36 BHCs exhibited Rank $>$6.2, with a $>$5$\sigma$ difference from NS LMXB spectra. 

\begin{centering}
\begin{figure}[!ht]
\epsscale{1.1}
\plotone{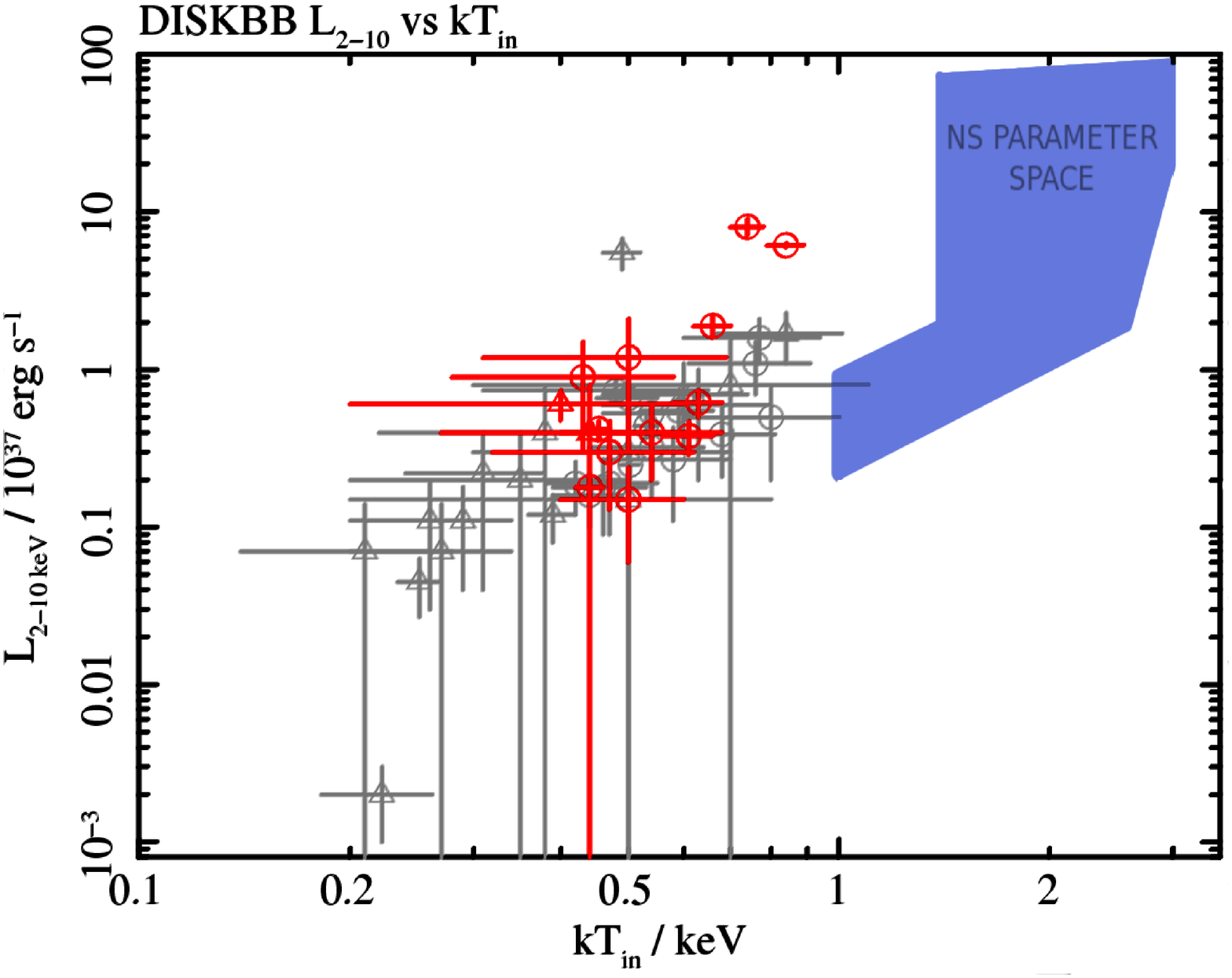}
\caption{Disk blackbody temperature vs. 2--10 keV luminosity for best fit double thermal models to 46 out of 50 BHCs; we were unable to fit double thermal models to the other 4 BHCs. Open circles represent persistent X-ray sources, while open triangles represent transients. Red symbols indicate BHCs in globular clusters. The shaded region indicates the NS LMXB parameter space \citep{barnard13}.}\label{tldbb}
\end{figure}
\end{centering}

  Figure~\ref{tldbb} shows $kT_{\rm DBB}$ vs. $L_{\rm DBB}$ for 46 BHCs; the double thermal model was unconstrained for 4 BHCs. Circles represent persistent X-ray sources, while triangles represent transients; red symbols indicate BHCs in globular clusters. None of our BHCs had best fits inside the NS LMXB region of $kT_{\rm DBB}$ vs. $L_{\rm DBB}$ parameter space, although some BHCs were consistent within 3$\sigma$. We see a natural systematic  correlation between temperature and 2--10 keV luminosity: lower temperature emitters contribute less to the 2--10 keV flux. The transients tended towards lower temperatures than the persistent sources; this is consistent with the transients exhibiting thermally dominated states rather than hard states \citep{remillard06}. Of the 15 GC BHCs in our sample, 13 differ from NS LMXB spectra at confidence levels $>$3$\sigma$ (2 transients, 11 persistent X-ray sources); 9 differ from NS LMXBS spectra at $>$5$\sigma$ confidence levels. These GC BHCs are particularly interesting because there are no confirmed  GC BH XBs in our Galaxy, and there are no known persistent GC BHCs anywhere outside M31.


\begin{table*}[!t]
\begin{center}
\caption{Best fit double thermal models to our BHCs, and  comparisons with the Galactic NS XBs. We first provide the name of the BHC, following \citet{barnard13}; we additionally include B045 and B375, which are outside the region covered by our Chandra survey. We then give the observation with the best spectrum. Next we provide the temperatures and luminosities of the disk blackbody and blackbody components, along with the disk blackbody contribution to the total flux. Finally, we give the Rank (i.e. $-$log[$P_{\rm NS}$]), class of BHC (strong or plausible), and comments. These comments indicate transients (T),  ULXs (U), sources that have been promoted from plausible to strong BHCs (P$>$S), and BHCs residing in globular clusters (G); the name of the cluster is given in parentheses. ``Soft'' indicates a spectrum that has little flux above 2 keV. Numbers in parentheses indicate 1$\sigma$ uncertainties in the last digit.} \label{sumtab}
\renewcommand{\tabcolsep}{3pt}
\renewcommand{\arraystretch}{1.2}
\begin{tabular}{cccccccccccc}
\tableline\tableline
BHC&  $kT_{\rm DBB}$ / keV & $L_{\rm DBB}^{2-10}$/10$^{37}$ & $kT{\rm BB}$ / keV & $L_{\rm BB}^{2-10}$/10$^{37}$ & $f_{\rm DBB}$ & Rank & Class & Comments\\
\tableline 
B045 & 0.66 $\pm$ 0.04 & 1.9 $\pm$ 0.3 & 1.51 $\pm$ 0.06 & 9.8 $\pm$ 0.3 & 0.16 $\pm$ 0.03 & 68 & S & G (B045) \\
B375 & 0.74 $\pm$ 0.04 & 8.0 $\pm$ 0.9 & 1.48 $\pm$ 0.06 & 23.0 $\pm$ 0.9 & 0.41 $\pm$ 0.05 & 30 & S & G (B375) \\
S109 & 0.5 $\pm$ 0.19 & 1.2 $\pm$ 0.9 & 1.3 $\pm$ 0.3 & 7.2 $\pm$ 1.1 & 0.14 $\pm$ 0.11 & 4.7 & S & G (B082) P$>$S \\
S111 & 0.7 $\pm$ 0.4 & 0.8 $\pm$ 0.8 & 0.89 $\pm$ 0.12 & 5.1 $\pm$ 0.9 & 0.14 $\pm$ 0.14 & 8.41 & S & T \\
S117 & $\dots$   & $\dots$   & $\dots$   & $\dots$   &    &  & S & T Soft ($<$2 kEV) \\
S122 & 0.43 $\pm$ 0.15 & 0.90 $\pm$ 0.60 & 0.9 $\pm$ 0.6 & 7.1 $\pm$ 1 & 0.11 $\pm$ 0.077 & 9 & S & G (B086) P$>$S \\
S151 & 0.5 $\pm$ 0.05 & 0.66 $\pm$ 0.19 & 1.2 $\pm$ 0.07 & 6.9 $\pm$ 0.3 & 0.09 $\pm$ 0.03 & 73 & S &  \\
S159 & 0.54 $\pm$ 0.14 & 0.4 $\pm$ 0.2 & 1.3 $\pm$ 0.2 & 2.3 $\pm$ 0.18 & 0.27 $\pm$ 0.1 & 4.2 & S & G (B096) P$>$S \\
S167 & 0.8 $\pm$ 0.2 & 0.5 $\pm$ 0.3 & 1.9 $\pm$ 0.6 & 2 $\pm$ 0.2 & 0.20 $\pm$ 0.12 & 1.9 & P &  \\
S168 & 0.53 $\pm$ 0.02 & 0.44 $\pm$ 0.07 & 1.38 $\pm$ 0.08 & 3.27 $\pm$ 0.12 & 0.12 $\pm$ 0.02 & 187 & S &  \\
S179 & 0.503 $\pm$ 0.014 & 0.51 $\pm$ 0.05 & 1.47 $\pm$ 0.05 & 3.98 $\pm$ 0.11 & 0.11 $\pm$ 0.01 & $>$320 & S &  \\
S199 & 0.44 $\pm$ 0.17 & 0.4 $\pm$ 0.4 & 1.3 $\pm$ 0.30 & 7.8 $\pm$ 1.7 & 0.05 $\pm$ 0.05 & 18 & S & T G (B117) P$>$S \\
S214 & 0.390 $\pm$ 0.030 & 0.120 $\pm$ 0.040 & 0.940 $\pm$ 0.060 & 2.039 $\pm$ 0.110 & 0.056 $\pm$ 0.019 & 208.473 & S & T  \\
S223 & 0.48 $\pm$ 0.17 & 0.74 $\pm$ 0.11 & 1.9 $\pm$ 0.5 & 1.7 $\pm$ 0.3 & 0.30 $\pm$ 0.1 & 4.5 & S &  \\
S233 & 0.35 $\pm$ 0.15 & 0.2 $\pm$ 0.2 & 0.84 $\pm$ 0.18 & 3 $\pm$ 0.5 & 0.06 $\pm$ 0.06 & 17 & S & T \\
S236 & 0.60 $\pm$ 0.14 & 0.7 $\pm$ 0.4 & 1.5 $\pm$ 0.3 & 5 $\pm$ 0.6 & 0.12 $\pm$ 0.07 & 7.64 & S & T P$>$S \\
S251 & 0.49 $\pm$ 0.03 & 5.5 $\pm$ 1.2 & 0.67 $\pm$ 0.08 & 2.8 $\pm$ 1.2 & 0.66 $\pm$ 0.20 & 147 & S & T U \\
S265 & 0.59 $\pm$ 0.07 & 0.55 $\pm$ 0.16 & 1.3 $\pm$ 0.2 & 1.71 $\pm$ 0.16 & 0.24 $\pm$ 0.07 & 11.1 & S &  \\
S269 & 0.68 $\pm$ 0.13 & 0.39 $\pm$ 0.18 & 1.9 $\pm$ 0.6 & 1.6 $\pm$ 0.3 & 0.20 $\pm$ 0.1 & 3.7 & S &  \\
S276 & $\dots$   & $\dots$   & $\dots$   & $\dots$   & 0.06   &  & P & T \\
S286 & 0.58 $\pm$ 0.12 & 0.27 $\pm$ 0.16 & 1.2 $\pm$ 0.19 & 1.95 $\pm$ 0.18 & 0.12 $\pm$ 0.07 & 9 & S & P$>$S \\
S287 & 0.251 $\pm$ 0.017 & 0.045 $\pm$ 0.018 & 0.46 $\pm$ 0.04 & 0.56 $\pm$ 0.05 & 0.074 $\pm$ 0.030 & $>$320 & S & T \\
S289 & 0.76 $\pm$ 0.15 & 1.1 $\pm$ 0.4 & 1.6 $\pm$ 0.2 & 4.8 $\pm$ 0.4 & 0.19 $\pm$ 0.07 & 4.7 & S & P$>$S \\
S293 & 0.4 $\pm$ 0.2 & 0.61 $\pm$ 0.13 & 1.50 $\pm$ 0.3 & 2.30 $\pm$ 0.2 & 0.21 $\pm$ 0.05 & 9 & S & T G (B128) \\
S297 & 0.44 $\pm$ 0.05 & 0.16 $\pm$ 0.06 & 0.99 $\pm$ 0.1 & 1.12 $\pm$ 0.09 & 0.13 $\pm$ 0.05 & 46 & S &  \\
S299 & 0.42 $\pm$ 0.04 & 0.19 $\pm$ 0.07 & 1.06 $\pm$ 0.05 & 4.23 $\pm$ 0.18 & 0.043 $\pm$ 0.016 & 208 & S &  \\
S300 & 0.84 $\pm$ 0.17 & 1.7 $\pm$ 0.6 & 2.3 $\pm$ 1 & 2.9 $\pm$ 0.4 & 0.37 $\pm$ 0.14 & 0.5 & P & T \\
S322 & 0.38 $\pm$ 0.16 & 0.4 $\pm$ 0.4 & 1.7 $\pm$ 1.2 & 4 $\pm$ 2 & 0.09 $\pm$ 0.10 & 7 & S & T \\
S327 & 0.83 $\pm$ 0.09 & 6.6 $\pm$ 1.5 & 1.62 $\pm$ 0.16 & 16.9 $\pm$ 0.7 & 0.28 $\pm$ 0.07 & 6.4 & S & G (B135) \\
S330 & 0.222 $\pm$ 0.04 & 2.00E-003 $\pm$ 1.00E-003 & 0.76 $\pm$ 0.1 & 0.24 $\pm$ 0.03 & 0.008 $\pm$ 0.004 & $>$320 & S & T NS HS? \\
S331 & 0.29 $\pm$ 0.05 & 0.11 $\pm$ 0.07 & 0.7 $\pm$ 0.2 & 0.38 $\pm$ 0.09 & 0.22 $\pm$ 0.152 & 51 & S & T \\
S335 & 0.77 $\pm$ 0.17 & 1.6 $\pm$ 0.5 & 2.10 $\pm$ 0.4 & 4.2 $\pm$ 0.6 & 0.28 $\pm$ 0.09 & 2.0 & P & \\
S339 & 0.31 $\pm$ 0.07 & 0.22 $\pm$ 0.18 & 0.89 $\pm$ 0.05 & 14.1 $\pm$ 1 & 0.015 $\pm$ 0.013 & 315 & S & T U \\
S345 & 0.54 $\pm$ 0.08 & 0.53 $\pm$ 0.08 & 1.31 $\pm$ 0.07 & 3.35 $\pm$ 0.13 & 0.14 $\pm$ 0.02 & 58 & S &  \\
S353 & 0.5 $\pm$ 0.3 & 0.15 $\pm$ 0.15 & 1.0 $\pm$ 0.17 & 1.4 $\pm$ 0.2 & 0.10 $\pm$ 0.1 & 6.8 & S & T \\
S358 & 0.46 $\pm$ 0.07 & 0.18 $\pm$ 0.09 & 1.01 $\pm$ 0.08 & 2.66 $\pm$ 0.16 & 0.06 $\pm$ 0.03 & 47 & S &  \\
S365 & 0.27 $\pm$ 0.07 & 0.07 $\pm$ 0.07 & 0.65 $\pm$ 0.06 & 2.4 $\pm$ 0.3 & 0.028 $\pm$ 0.029 & 118 & S & T \\
S372 & 0.44 $\pm$ 0.02 & 0.18 $\pm$ 0.03 & 1.13 $\pm$ 0.06 & 1.96 $\pm$ 0.09 & 0.084 $\pm$ 0.015 & 321 & S & G (B143) \\
S373 & 0.61 $\pm$ 0.05 & 0.38 $\pm$ 0.09 & 1.61 $\pm$ 0.16 & 2.72 $\pm$ 0.15 & 0.12 $\pm$ 0.03 & 41 & S & G (B144) \\
S386 & 0.50 $\pm$ 0.10 & 0.15 $\pm$ 0.09 & 0.96 $\pm$ 0.07 & 1.63 $\pm$ 0.12 & 0.08 $\pm$ 0.05 & 18.3 & S & G (B146) P$>$S \\
S389 & 0.47 $\pm$ 0.08 & 0.19 $\pm$ 0.1 & 0.97 $\pm$ 0.1 & 1.51 $\pm$ 0.13 & 0.11 $\pm$ 0.06 & 25 & S &  \\
S391 & 0.47 $\pm$ 0.15 & 0.30 $\pm$ 0.17 & 1.06 $\pm$ 0.16 & 2.0 $\pm$ 0.3 & 0.13 $\pm$ 0.08 & 10.1 & S & G (B148) P$>$S \\
S396 & 0.26 $\pm$ 0.06 & 0.11 $\pm$ 0.08 & 0.98 $\pm$ 0.05 & 15.5 $\pm$ 0.9 & 0.007 $\pm$ 0.005 & $>$320 & S & T \\
S411 & $\dots$   & $\dots$   & $\dots$   & $\dots$   &    &  & P & T \\
S415 & 0.453 $\pm$ 0.014 & 0.42 $\pm$ 0.05 & 1.36 $\pm$ 0.05 & 5.11 $\pm$ 0.15 & 0.08 $\pm$ 0.01 & $>$320 & S & G (B153) \\
S438 & $\dots$   & $\dots$   & $\dots$   & $\dots$   &    &  & P & T G(B163) \\
S448 & 0.21 $\pm$ 0.07 & 0.07 $\pm$ 0.07 & 0.41 $\pm$ 0.02 & 4.1 $\pm$ 0.4 & 0.02 $\pm$ 0.02 & $>$320 & S & T \\
S484 & 0.54 $\pm$ 0.1 & 0.32 $\pm$ 0.17 & 1.3 $\pm$ 0.2 & 1.8 $\pm$ 0.18 & 0.15 $\pm$ 0.08 & 9.4 & S &  \\
S487 & 0.63 $\pm$ 0.05 & 0.62 $\pm$ 0.12 & 1.6 $\pm$ 0.2 & 3.1 $\pm$ 0.2 & 0.17 $\pm$ 0.03 & 29 & S & G (B185) \\
S497 & 0.63 $\pm$ 0.17 & 0.6 $\pm$ 0.4 & 1.6 $\pm$ 0.4 & 2.7 $\pm$ 0.3 & 0.18 $\pm$ 0.12 & 3.0 & S & T P$>$S \\

\tableline
\end{tabular}

\end{center}
\end{table*}

\subsubsection{Comparison with a Bright  NS XB in M31}
 RX J0042.6+4115 is usually the brightest X-ray source in M31 (0.3--10 keV luminosity $\sim$5--6$\times 10^{38}$ erg s$^{-1}$), and was classified as a Z-source (NS LMXB) after exhibiting an apparently tri-modal color-intensity diagram \citep{barnard03b}.  We decided to model the X-ray emission  with double thermal models to see if it was  consistent with the findings of Lin et al. (2007, 2009, 2010, 2012).

The highest quality spectrum for RX J0042.6+4115 came from the 2002 January 6 XMM-Newton observation 0112570101 (PI M. Watson). The total exposure time was $\sim$60 ks, and we ignored the first 10 ks due to an unstable background. The resulting spectrum contained $\sim$30000 net source counts. During this observation, RX J0042.6+4115 appears to have been in a Cyg-like horizontal branch state \citep{barnard03b}.
We were unable to obtain successful fits to the spectrum with a double thermal emission model; this is expected, since Lin et al (2009, 2012) required a disk blackbody + blackbody + Comptonization emission model for the horizontal branch. This model was able to successfully describe the RX J0042.6+4115 spectrum with parameters consistent with Lin et al. (2009): $kT_{\rm DBB}$ $\sim$1.9 keV, $kT_{\rm BB}$ $\sim$2.5 keV, and $\Gamma$ $\sim$2.4, $\chi^2$/dof = 738/705. However, we were unable to constrain the parameters,  despite the high quality of the data; this is probably because the blackbody component  peaks at $\sim$3 kT, i.e. $\sim$7 keV, near the upper limit to the XMM-Newton energy band.

\subsection{Estimating the BH population within 6$'$ of M31*}

In Barnard et al. (2014a), we found that the number of sources consistent with AGN in our Chandra survey of sources within 20$'$ of M31*  was well below the number predicted from the 0.5--10 keV AGN flux distribution obtained by \citet{georgakakis08}. However, when we restricted our sample to those within 6$'$ of M31* we saw a surplus. Hence, the observed deficit is  dominated by instrumental effects. With this in mind, we decided to estimate the black hole contribution to the X-ray population within 6$'$ of M31*, from the duty cycles of transients within this region. 

Our survey contains 216 X-ray sources within 6$'$ of M31*, of which 126 are probably XBs, 66 are consistent with AGN, and 22 are known stars, novae etc; the 0.3--10 keV detection limit is $\sim10^{35}$ erg s$^{-1}$, although it is not complete at this level. The 126 probable XBs include 33 X-ray transients. We found 34 of our BHCs in this region, 20 persistent, and 14 transient. To date, we have no information on the accretors in the other 19 transients; they could contain black holes or neutron stars. 

We estimated the maximum duty cycle for the unclassified transients from the intervals when transient was not detected at the 3$\sigma$ level (i.e. like  DC2 for our transient BHCs in Table 1). The mean maximum  duty cycle for transients $>$10$^{35}$ erg s$^{-1}$ was 0.13; this would suggest a total of 254 transients within 6$'$ of M31*, 108 of which containing BHs. As a result, we expect $>$40\% of XBs within 6$'$ of M31 to contain BHs.  Assuming the median duty cycle for the transients within 6$'$ of M31* (0.07) yielded essentially the same results. The BH fraction would be higher if the actual duty cycle was smaller, or if some of the unclassified transients contain BHs.

Similarly, we observed 24 probable XBs which exceeded 10$^{38}$ erg s$^{-1}$ in the 0.3--10 keV band; 20 of these were BHCs. This sample included 11 transients (10 BHC transients), with a mean maximum duty cycle  0.09. These findings suggest that $>$90\% of sources that exceed 10$^{38}$ erg s$^{-1}$ within 6$'$ of M31* contain black holes. By contrast, the majority of Milky Way (MW) XBs $>$10$^{38}$ erg s$^{-1}$ are NS XBs \citep{grimm02}.

The 2007 MW X-ray binary catalog \citep{liu07} contains 103 X-ray transients, of which 83 are classified with NS or BHC accretors. BHCs represent $\sim$50\% of the total  MW  transient population with known accretors, but $\sim$70\% of transients with known distances and luminosities $>$10$^{37}$ erg s$^{-1}$. We found that 31 of the 33 transients within 6$'$ of M31 exceeded 10$^{37}$ erg s$^{-1}$ at some point during our 13 year survey; if $\sim$70\% of these transients contain BHCs, then $>$60\% of the XBs within 6$'$ of M31* could contain BHCs.

\section{Discussion}
In this work, we expand upon  \citet{barnard13} where we compared the spectra of 35 BHCs with the full range of  neutron star spectra. \citet{lin09} have applied a double thermal emission model to a transient Z source that exhibited all types of NS LMXB behavior; this model gave good fits except for hard state spectra, and horizontal branch spectra (where a Comptonized component is requred, which dominates hard state spectra). \citet{lin07} first presented this model as a NS analog to the BH thermally dominated state; the main strength of the model when applied to the two original tranisents was that luminosity followed $T^4$ for both thermal components (Lin et al., 2007). However, the temporal and spectral evolution of high inclination LMXBs indicates a substantial extended comptonized component in LMXBs throughout the luminosity range \citep[and references within]{church04}. Nevertheless, the work of Lin et al. (2007, 2009, 2012) has been extremely useful because it allows us examine the gamut of NS LMXB behavior in a single parameter space. 

BH LMXBs exhibit a thermally dominated state that has never been observed in NS LMXBs; this state is usually observed in X-ray transients \citep{remillard06}. We examined $\sim$50 X-ray transients identified in our Chandra survey \citep{barnard14a}, and found 13 suitable for spectral fitting. The remaining transients are possible BHCs, but may also contain NS accretors; further observations may clarify the identities of these systems. We used an improved method for comparing our BHC spectra with NS LMXB spectra for these 13 transients, our 35 original BHCs, and the GC BHCs B045  and B375 for a total of 50 BHCs. We found that 42 exhibited spectra that differed from the NS spectra at a $>$3$\sigma$ level, and 36 at a $>$5$\sigma$ level; these were all classed as strong BHCs, except for S330 which exhibited a luminosity consistent with a NS XB hard state. The spectrum of S117 was unable to constrain the double thermal model, but was too soft to be a NS XB, and S117 is also considered a strong BHC. The remaining sources were classed as plausible BHCs; 10 BHC that were previously identified as plausible in \citet{barnard13} were promoted to strong BHCs. We expected hard state and thermally dominated BHCs to be inconsistent with NS spectra. However, we also found some steep power law spectra that were inconsistent with NS spectra; these were particularly soft, with $\Gamma$ $\ga$3. We certainly do not infer that all BH spectra should be separable from NS spectra.

Using this method, we may identify BHCs in many galaxies, including our own. The known Galactic BH LMXBs are all transient, or turned on recently \citep[see][and references within]{remillard06}. This is because they were identified with a method that requires observations of optical emission lines in the donor spectrum;  however, the optical emission of persistently bright X-ray sources is dominated by reprocessed X-rays from the disk \citep{vp94}. Our X-ray method has no such biases, and may reveal further BHCs in the known Galactic LMXB population.

We also  examined a bright M31 XB ($>$10$^{38}$ erg s$^{-1}$) that is expected to contain a NS accretor (S209 Barnard et al. 2014a). A double thermal model fit  yielded parameters consistent with the NS systems studied by Lin et al. (2007, 2009, 2012). Hence, the observed differences between our BHCs and the Galactic NS XBs studied by Lin et al. (2007, 2009, 2012) is not due to differences between the RXTE, Chandra, and XMM-Newton observatories.  

We have identified 126 probable X-ray binaries within 6$'$ of M31* \citep{barnard14a}, of which 34 are BHCs; 33 of these systems are transient, including 14 BHCs. The mean maximum duty cycle of the transient systems was 0.13, suggesting that $>40$\% of XBs within 6$'$ of M31* contain BHs. However, our results suggest that BH XBs contribute 90\% of XBs exceeding 10$^{38}$ erg s$^{-1}$ in this region. This result provides further substantial difference in the evolution histories of M31 and the Milky Way, since the majority of MW X-ray sources exceeding 10$^{38}$ erg s$^{-1}$ are NS XBs \citep{grimm02}. We already know that M31 contains $\sim$30 times as many bright GC X-ray sources ($>$10$^{37}$ erg s$^{-1}$) as the MW \citep{barnard14a}, and could contain $\sim$4--5 times as many XBs over all \citep{stiele11,barnard14a}.

\section*{Acknowledgments}
 We thank the anonymous refereee for useful suggestions that considerably improved this paper. This research
has made use of data obtained from the {\em Chandra} satellite,
and software provided by the {\em Chandra} X-Ray Center (CXC).
We also include analysis of data from {\em XMM-Newton}, an ESA
science mission with instruments and contributions directly
funded by ESA member states and the US (NASA); we are very grateful to Norbert Schartel and the XMM-Newton team for granting our TOO observation. R.B. is
funded by {\em Chandra} grants GO2-13106X and GO1-12109X,
along with {\em HST} grants GO-11833 and GO-12014.





{\it Facilities:} \facility{\em CXO (ACIS)} \facility{XMM-Newton (pn)} 








\end{document}